\documentclass[preprint]{article}
\usepackage{graphicx}
\usepackage{amsmath}

\newcommand{\be}{\begin{equation}}
\newcommand{\ee}{\end{equation}}
\newcommand{\ba}{\begin{eqnarray}}
\newcommand{\ea}{\end{eqnarray}}

\begin{document}
\hoffset=-.4truein\voffset=-0.5truein
\setlength{\textheight}{8.5 in}
\begin{titlepage}
\begin{center}
\vskip 40 mm
{\large \bf Knots  from a random matrix theory with  replicas
\footnote{ This  article is a contribution to  the Michael Fisher book for his memory, edited by A. Aharony, 
O. Entin-Wohlman, D. Huse and L. Radzihovsky.  }
}
\vskip 10mm
{\bf Shinobu Hikami}
\vskip 5mm

{Okinawa Institute of Science and Technology Graduate University,\\
 1919-1 Tancha, Okinawa 904-0495, Japan.\\
e-mail: hikami@oist.jp}

\vskip 5mm
{\bf Abstract}
\vskip 3mm
\end{center}
A classical knot is described by a one-stroke trajectory of a string with entanglements. The replica method is  a powerful tool in statistical mechanics for dealing with string-like objects like polymers or self-avoiding walks. We consider here the ($N \to 0$) replica limit for Gaussian means of products of traces of $N \times N$ Hermitian matrices, which correspond to one-stroke graphs for knots. The Seifert surfaces of knots and links are thus related to a random matrix model. The zeros of Alexander polynomials on the unit circle are discussed for the case of n-vertices in analogy with the Yang-Lee edge singularity. The extension of one-matrix models to higher dimensional knots is considered, and also to the half-integral level $k$ in a Chern-Simons gauge theory.



 \end{titlepage}
 \vskip 3mm

\section{Introduction}
\vskip 2mm

    The replica method is  a powerful tool in statistical mechanics to deal with disordered systems. The critical behavior of a single polymer, or equivalently the geometric properties of self-avoiding walks  are given by a zero-replica limit ($N \to 0$) of the $O(N)$ vector model \cite{DeGennes}, allowing the use of 
   Wilson-Fisher $\epsilon$ expansion \cite{WilsonFisher}.  The Anderson localization in a random potential is  known  to be   the zero replica limit of a  Grassmannian nonlinear $\sigma$ model, opening the way to  a renormalization group analysis \cite{Hikami}. The intersection numbers of $p$-spin curves for the moduli space of Riemann surfaces  \cite{Kontsevich,Witten1} 
have also been  investigated
  by a replica method using  a random matrix theory with an external source \cite{BrezinHikami1,BrezinHikami2,BrezinHikami3}. The replica limits ($N \to 0$) appears  as  one stroke trajectories for Wilson lines in a quantum field theory.
  
 Generally  those  problems are related to one-stroke trajectories of  strings.  We shall examine in this article how the replica method for a Gaussian random matrix theory
 corresponds to  the theory of knots studied intensively in mathematics and applied in statistical mechanics (for instance, see the text book of knots \cite{Murasugi}). This classical theory
  has bee extended to higher dimensional knots, such as 2-knots \cite{Kamada},  which may also be put in correspondence with matrix models.

    \vskip 3mm     
\section{Knot graphs and matrix models}
\vskip 2mm
  Let us recall first some earlier results on Gaussian matrix integrals.   We consider  Gaussian averages as in \cite{BrezinHikami3}, of the form
 \be\label{average}
 <X(M)> = \frac{1}{Z} \int dM X(M) e^{-\frac{1}{2} {\rm tr} M^2}
 \ee
 where $X(M)$ is  a product of multi-traces of ${\rm tr} M^n$, i.e. $(\prod_i {\rm tr} M^{n_i})$. The matrices $M$ are Hermitian $N\times N$  and $Z$ is a normalization, which ensures that $<1>=1$. 
 
 For multi-traces such as $<\frac{1}{N} \rm tr (M^{n_1} \cdots {\rm tr} M^{n_k})>$ the application of Wick's theorem  produces a sum of Feynman graphs with double lines between the $k$ vertices.  
  In the $N\to 0$ limit the only graphs which remain consist of a single stroke line going along the propagators from vertex to vertex.  From our previous work on intersection  numbers 
 of the moduli space of curves on a Riemann surface   \cite{BrezinHikami1}, starting with the Konsevitch model \cite{Kontsevich},  we have found an explicit formula for the replica limit of the generating function of multi traces . If 
 \be\label{replica}
U(\sigma_1,...,\sigma_n) = \frac{1}{N} < \prod_{i=1} ^n  {\rm tr} e^{\sigma_i M}>
\ee
denotes this generating function, 
its zero-replica limit is given by 
 \be\label{rf}
 \lim_{N\to 0} U(\sigma_1,...,\sigma_n) = \frac{2^n}{\chi^2} \prod_{i=1} ^n  {\rm sinh} \biggl( \frac{\chi \sigma_i}{2}\biggr)
 \ee
 with $\chi= \sum_{i=1}^n \sigma_i$.
 The expansion of the generating function (\ref{rf}) in powers of the $\sigma_i$  leads easily to explicit results such as 
selection rules, for instance,
 \be
 \lim_{N\to 0} \frac{1}{N} < ({\rm tr} M^3)^k> = 0
 \ee
  unless $k=2 $ ( mod 4).  For other powers (\ref{average}) one finds for instance
  \be\label{select}
  \lim_{N\to 0} \frac{1}{N} < ({\rm tr} M^3)^{4g-2}> = \frac{3^{3g-2}2^{-2g} (6g-4)!(4g-2)!}{g! (3g-2)!} .
  \ee
 In our earlier work  $g$ was identified with the genus of the Riemann surface whose intersection was the r.h.s. of (\ref{select}). 
 The Gaussian mean of above case in the replica limit has a number, which counts the different contraction of legs. For instance, $g=1$ case, this number becomes 3, which is
 3 possibilities of contraction of 6 legs of the two vertices in the replica limit. Thus such numbers are related to the numbers of different knots and circles.
  
  On the other hand mathematicians describe knots with the help of Seifert surfaces,
 whose edge  provides a directed knot \cite{Murasugi}.
 A Seifert surface is made of disks connected each to one another by  twisted bands (Seifert band). One  establishes a correspondence between such a surface  with a  Feynman diagrams, in which a disk corresponds to a vertex, and a connected Seifert band between disks  as a propagator.   
  From the point of view of knots, this replica limit  in the matrix  side involves both knots and unknotted circles : the correspondence between knots and matrix averages is not one-to-one.
 The matrix model ignores  the two ways of the over or down trails in the Seifert band. The matrix model belongs to two dimensions, while
 knots belong to three dimensions, embedded into $S^3$. The description of  directed trails over and down 
 is  characterized by a skein relation \cite{Murasugi}.
 If one endows the vertex of a matrix graph with a sign, like the head and tail of a coin, the connection of two vertices (head and tail) produces a Seifert band with the  two choices of up or down crossing.

 The  standard  catalogue of knots (Rolfsen catalogue)  lists  the knots with their signature such as $3_1, 4_1, 5_1,  5_2, ...$. The knot of $3_1$,  named  trefoil, $4_1$ is a figure 8 knot etc.  The correspondence with matrix averages in the zero-replica limit is built as follows. For instance, a trefoil knot ($3_1$) corresponds to $\lim_{N \to 0} \frac{1}{N}<{(\rm tr} M^3)^2>$ in (\ref{average}) .  The trivalent vertex is described by ${\rm tr} M^3$. 
 The average is Gaussian average, but for the Seifert surface,   the Seifert band is taken as a connector (propagator). It is obtained by flipping half of the vertices, and connect all
 vertices by Seifert bands.
 
 
 
 For the  $5_1$ knot, the corresponding Gaussian average is  $<({\rm tr} M^5)^2>$. (From now on a factor $1/N$ and a limit $N \to  0$ is meant). The Gaussian average of this is 165, which is the sum of three different graphs, and the sum of the numbers of the contractions of these graphs becomes 165. The knot of $5_1$ has a contraction number as 5. These different graphs lead to  $5_1$ and circles. For the $7_1$ knot, the corresponding term is $<(\rm tr M^7)^2>$. In general, the term of $<({\rm tr} M^n)^2>$ 
 gives the knot, named as $n_1$, where $n$ is odd.
 
 The $4_1$ (figure-8) knot corresponds to the average $<({\rm tr} M^4)({\rm tr} M^2)^2>$.  It is easy to find the Gaussian means of the product of matrices for each Seifert graphs of knots, by the trails which take the opposite direction path at the crossing points of knot graphs.  Gaussian means of (\ref{average}) include unknotted circles, therefore we need some selection
 rule which separate knots graphs from unknotted circles. The separation of the unknotted circles is made through a skein relation \cite{Murasugi}. The following table lists the correspondence between the Rolfsen numbers and the Gaussian averages. 
  \vskip 3mm
  {\bf Table A}: Standard knot notation (left) and Replica limit of Gaussian mean (right) with Seifert band contractions denoted by $s$. 
  \vskip 2mm
  \begin{align*}
& 3_1 \hskip 5mm <({\rm tr}M^3)^2>_s \nonumber\\
  & 4_1 \hskip 5mm <({\rm tr} M^4) ({\rm tr} M^2)^2>_s \nonumber\\
  & 5_1 \hskip 5mm <({\rm tr} M^5)^2>_s\nonumber\\
  & 5_2 \hskip 5mm < ({\rm tr} M^3)^2 ({\rm tr} M^2)^2>_s \nonumber\\
  & 6_1 \hskip 5mm < ({\rm tr} M^4) ( {\rm tr} M^2)^4>_s\nonumber\\
  & 6_2 \hskip 5mm <({\rm tr} M^6)({\rm tr} M^4) ({\rm tr} M^2)>_s\nonumber\\
  & 6_3 \hskip 5mm <({\rm tr} M^6)({\rm tr} M^3)^2>_s \nonumber\\
  & 7_1 \hskip 5mm < ({\rm tr} M^7)^2>_s\nonumber\\
  & 7_2 \hskip 5mm <({\rm tr} M^3)^2 ({\rm tr} M^2)^4>_s \nonumber\\
  & 7_3 \hskip 5mm <({\rm tr} M^5)^2 ({\rm tr} M^2)^2>_s  \nonumber\\
  & 7_4 \hskip 5mm <({\rm tr} M^3)^2 ({\rm tr} M^2)^4>_s \nonumber\\
  & 7_5 \hskip 5mm <({\rm tr} M^5)({\rm tr}M^4) ({\rm tr} M^3)({\rm tr}M^2)>_s  \nonumber\\
  & 7_6 \hskip 5mm <({\rm tr} M^7)({\rm tr} M^3)({\rm tr}M^2)^2>_s \nonumber\\
  & 7_7 \hskip 5mm <({\rm tr} M^6)({\rm tr}M^3)^2({\rm tr} M^2)>_s \nonumber\\
 \end{align*}

   The same Gaussian means  appear for  different knots. The difference is due to the different crossing ways in a Seifert band. 
   For instance, the knots of  $8_9,8_{17},8_{18},8_{19},8_{20},8_{21}$ have same the same Gaussian means.  They correspond indeed to several different configurations
   of the crossings in   the Seifert band. The contractions of $8_9$, $8_{18}$ and $8_{17}$ are all different (Feynman diagrams are different). Starting from the alternating knot of $8_{17}$, by the changes of the paired crossing of over or down of these Seifert bands, other knots of non-alternating  knots $8_{19}$, $8_{20}$ and $8_{21}$ are obtained.
       
   Some cases appear with the same Gaussian mean but different Seifert graphs. The cases of $7_2$ and $7_4$ have the same Gaussian mean $<({\rm tr} M^3)^2({\rm tr }M^2)^4>_s$, but
   the Seifert graphs are different. 
   
   The Gaussian mean is evaluated by Wick contraction, which is a commutative algebraic procedure. However, if one writes a Feynman diagram in the zero-replica limit, using for the drawing double line propagators, the order chosen for writing the diagram matters. According to the  order chosen for the double line propagators, sometimes one generates knotted diagrams, but some ordering of lines may lead to diferrent knots or even  unknotted circles. For instance
   in the case of $<({\rm tr} M^3)^2>$, the maximum crossing diagram gives the trefoil knot $3_1$ or a circle,  depending on the choices for drawing a maximum crossing Feyman diagram. This is
   the same in a Seifert band, which gives a knot ($3_1$) or a circle, depending upon the choices of which line goes over or down at a crossing.
   
   If we use a standard Feynman diagram, instead a twisted Seifert graphs, there appear several graphs which are different depending how we write the propagators in a time order
   on a paper, i.e. the way of the crossing of the ribbons. For instance in the case of $<({\rm tr} M^3)^2>$, the maximum crossing graphs give  one-stroke graphs. If the change  of order of 
   ribbons crossing, we have $3_1$ and also $4_1$. We have mentioned that the head and tail connected  vertex of Seifert surface  gives $3_1$ or a circle. Thus there is a difference
   between a standard Feynman diagram for matrices and twisted Seifert graphs. We shall discuss the way of the distinctions for different knots by the numbering in the subsequent section.
   
\vskip 3mm
\section{Chern-Simons gauge theory and random matrix}
\vskip 3mm   
  This Seifert band  is similar to the propagator  of following Chern-Simons action of gauge field \cite{Witten2}.
 
 \be\label{CS}
 \mathcal{L} = \frac{k}{4\pi}\int_{W}  {\rm tr} (A\wedge dA + \frac{2}{3} A\wedge A\wedge A )
 \ee
 where $k$ is integer, which corresponds to a level of Lie group, and $W$ is a three dimensional manifold.
 
 Near the crossing point of Seifert band, two edges are twisted, and it means the two edges are bound to each other. The two bounded edges are analogous to two bounded merons, which form an instanton.
 The merons were discussed in QCD \cite{Callan}, and they are also found in  a super conductor as a vortex, which  pairs condensate  into  two-dimensional superconductor \cite{Kosteritz, Tsuneto}. These
 bound states are described by an abelian or non-abelian gauge theory.
 
 The knot should be related to a Chern-Simons gauge theory, since the knot invariant polynomial (Jones polynomial) \cite{Jones} can be found from Chern-Simons gauge theory with a quantum parameter $q = e^{\frac{2\pi i}{p}}$ with $p=k+2$ \cite{Witten2}. This parameter $p$ has a meaning of a spin $p$, and $p$-spin curve on the Riemann surface has a topological invariant quantity of an intersection number of the moduli space \cite{Witten1, BrezinHikami3}
 
 The boundary of three dimensional Chern-Simons, which is turned to  two dimensions, described by Wess-Zumino-Witten (WZW) non-linear sigma model. Thus matrix model in the replica limit should be
 connected to WZW or Chern-Simons gauge theory.
 
 The Seifert surface is interpreted as a matrix model with head and tail two type vertices. The Seifert band shows the crossing point which has over or down two choices. This crossing can be interpreted as a twist operator of Pauli matrices $\sigma_x$ and $\sigma_y$ which are elements of $su(2)$ Lie algebra. The vertex of tail can be defined by the vertex  
 multiplied with twist operator $\sigma$. Then, the standard  Wick contraction with replica limit gives knots. The WZW model for $SU(2)$ Lie group naturally appears by taking this $su(2)$ Lie algebra with
 k-valence vertices where $k$ is a level of $SU(2)_k$. The knot polynomial of Jones can be evaluated from the skein relation with a quantum parameter $q= e^{\frac{2\pi i}{k+2}}$
 \cite{Witten2}. The twist operator, which is coupled to vertices of tail, is similar to the external source, which is deterministic in a random matrix theory \cite{BrezinHikami3}.
 
 In this respect, the super matrix $SU(n|m)$  \cite{BrezinHikami4s} may be useful. The replica limit is $n\to m$. When $n=2$, this supermatrix becomes $SU(2|2)$ and the previous Seifert band 
 with the twist operator $\sigma$ will appear. 
 
  \vskip 3mm
  \section{Numbering}
\vskip 2mm   
The correspondence between matrix models, in the zero-replica limit, and knots can be made one-to-one by numbering the edges.
We assign numbers to the edge-lines, numbering the floor to which this line belongs. The floors are connected to over or down floors by two types of Seifert bands ( staircase or escalator).
In the case of an over-connection from n-th to (n+1)-th floor, the number increases of one bit from $n$ to $n+1$. For a downward connection, $n$ becomes $n-1$. At the crossing point of a Seifert band, the total number is conserved. For incoming numbers $n$ and $m$,  the sum of outgoing  numbers is also $n+m$. 

For the trefoil knot, the edges of two vertices  carrying ${\rm tr} (M^3)$,  are assigned numberings  1 and 2. The Seifert bands band of this case is $1\to 2$ and $2 \to 1$, according to over and down twisted connections.
This assignment is consistent with the fact that the trefoil is an alternating knot. Along the directed path, the trefoil is written as the sequence of numbers  $(2121212)$, which is expressed as the paired sequence of the crossings

\be
 \left(
 \begin{array}{cccc}
2&1&2&1\\
1&2&1&2
 \end{array}
 \right)
 \ee
where 3, the sum of the numbers of the first and second row in the same column, is conserved . This matrix is a sequence of  2 by 2 scattering matrices.  Trivalent vertices can also provide  unknotted cases  if  one Seifert band can be changed from  over to down
and vice versa. Then we have
\be
 \left(
 \begin{array}{cccc}
2&3&2&1\\
1&0&1&2
 \end{array}
 \right)
 \ee
Although the number conservation, the sum of incoming and outgoing, holds,   the total sum  of  the first row is 8,  whereas  the sum  of the second row is 4. It shows that this case reduces to  an unknotted circle.

The number $n$ introduced here may be transformed to a phase $e^{i \theta n}$ with an angle $\theta$. The increase of the number $n$ to $n+1$ leads to a change $e^{i \theta n} \to
e^{i \theta (n+1)}$. This interpretation of phase is more close to a gauge field picture. This phase is related to parameter of Jones polynomial $t^n = e^{\frac{2i \pi n}{p}}$ if we take
$\theta$ as pure imaginary.

There are many non-alternating knots, which start from the eight crossing points of Rolsen table as $8_{19}, 8_{20}$ and $8_{21}$. The knot $8_{17}$ is an alternating knot. These four knots have same
graphs in the replica limit of matrix model, $<{\rm tr } M^8 ({\rm tr} M^4)^2>_s$. The non-alternating knots have a repeated connections of over or down crossings. Thus the numbers are
increasing or decreasing in several digits. So the numbers are needed to be greater than 2. 
For the alternating knot  of $8_{17}$, the numbers of sequence are $(1212121212121212)$, while the non-alternating knot $8_{19}$ has a sequence $(1232123212323212)$. These two
knots have the same graph except the over or down connections. The following graph represents the knots of $8_{17}, 8_{19}, 8_{20}$ and $8_{21}$. The three strands should be connected as the upper points to the lowest points. The path has upper to down direction. Then three directed circles are obtained which are connected by the 8 transverse rungs. The center strand which has 8 rungs represents ${\rm tr} (M^8)$, and other both sides column represents ${\rm tr} (M^4)$. The rung means the Seifert band, which has two choices of over or down directed path. For the case $8_{17}$, all segments of the center strand are assigned
by  2, and 1 for the vertical segments of  strands of both sides.  This alternating knot corresponds to the sequence $(1212121212121212)$. The graph is called traditionally as "Amida-kuji" (Kuji means a lottery). 
\vskip 3mm
\begin{center}\begin{tabular}{cccccc}
 & & & & & \\
  & \multicolumn{2}{|c|}{} & \multicolumn{2}{|c|}{} & \\\cline{4-5}
  & \multicolumn{2}{|c|}{} & \multicolumn{2}{|c|}{} & \\\cline{2-3}
  & \multicolumn{2}{|c|}{} & \multicolumn{2}{|c|}{} & \\\cline{4-5}
 & \multicolumn{2}{|c|}{} & \multicolumn{2}{|c|}{} & \\\cline{2-3}
 & \multicolumn{2}{|c|}{} & \multicolumn{2}{|c|}{} & \\\cline{4-5}
 & \multicolumn{2}{|c|}{} & \multicolumn{2}{|c|}{} & \\\cline{4-5}
 & \multicolumn{2}{|c|}{} & \multicolumn{2}{|c|}{} & \\\cline{2-3}
  & \multicolumn{2}{|c|}{} & \multicolumn{2}{|c|}{} & \\\cline{2-3}
 & \multicolumn{2}{|c|}{} & \multicolumn{2}{|c|}{} & \\
 \end{tabular}
\end{center}
\vskip 3mm

This numbering on amida-kuji can also distinguish the unknotted circle from the linked one. For the link of two circles, the path becomes $2\to 1 \to 2$ or $1\to 2\to1$. The starting number is same
as the ending number. For the unlinked circles, if they are place in an overlap configuration, the numbers become $1\to 2\to 3$. The starting number 1 becomes different from the ending
number 3. Thus this case is easily eliminated for the knots and links.

When the average of Gaussian mean is given, and the crossing points are $n$, it may be not the minimal number of the crossing points. The example is $<{(\rm tr} M^4)^2 {\rm tr} M^8>$,
which has 8 crossing points. By the choice of up and down crossing of 8 Seifert bands,  one reduces to knot $5_2$, which has 5 crossing points. This example appears in Teneva's transformation of a knot $5_2$ \cite{Saito}, and it has 4 coloring. The previous numbering of  floor-height becomes a sequence of $12101234323212321$, which involves a sequence of the increasing numbers $1234$. After Reidemeister transformation, it reduces to $5_2$ knot.

The numbering, which is shown here, is similar to introduction of the time. The increasing of the height of floor means the increasing of time. The standard treatment of knot invariants is based on 
Knitzhnik-Zamolodchikov equation with the Cauchy repeated integration \cite{Kontsevich}, which uses the integration of time. 
We find the one to one correspondence of the replica matrix model by the numberings to the different knot configurations. This may be same as Seifert surface, but the replica method
has an advantage for  generating all knot configurations. 

We consider the numbering of the height of floors. This can be generalized to any numbers such as a complex number of the root of unity. The general skein relation is \cite{Murasugi}
\be
x L_{+} - y L_{-} = w L_{0}
\ee
 where $x,y$ and $w$ are arbitrary.
One may take these numbers as $x=t^{-1}$, $y= t$ and $w= (\sqrt{t}- \frac{1}{\sqrt{t}})$, which leads to Jones polynomials \cite{Jones}. 

The skein relation is similar to the algebra of $su(2)$, which has Pauli matrices $\sigma_x,\sigma_y$ and $\sigma_z$ with $[\sigma_x,\sigma_y] = i \sigma_z$. The operator $\sigma_z$
corresponds to eraser of the rung of Amida-kuji, and $\sigma_x \sigma_y$ and $\sigma_y\sigma_x$ correspond to the over and down crossing of Seifert band. The introduction of the
operator $\hat \sigma$ for the vertices of the matrix model is alternative method to the numbering for the knot construction. This leads to  $SU(2)_k$ Wess-Zumino-Witten non-linear $\sigma$ model in two dimensions, which is
equivalent Chern-Simon gauge action in three dimensions.

The numbering of the segments of the vertices of the matrix in the replica limit gives   the Chern-Simons gauge field $A$ in (\ref{CS}). The $n$ point function $U(\sigma_1,...,\sigma_n)$ 
in (\ref{replica}) is similar to the multi-loop invariants of Chern-Simons theory,
\be
\hat\mu(\sigma_i,...\sigma_n) = \int_A \prod_{i=1}^n {\rm tr} ( e^{\int_{\sigma_i} A}) d\mu(A)
\ee
The matrix $M$ with a numbering turns to  a gauge invariant field, which has $SU(2)_k$ symmetry. The physical interpretation may be that the one-stroke trajectory knot is a string
(vortex flux ring) which is  a three-dimensional gauge invariant field.

\section{Derivation of replica formula}
\vskip 2mm     
 For the case of two vertices, we have from (\ref{rf})
\ba
&&\lim_{N\to 0} U(\sigma_1,\sigma_2) = \sigma_1\sigma_2 + \frac{1}{3}( \sigma_1^5 \sigma_2 + 2 \sigma_1^4 \sigma_2^2+ 2 \sigma_1^3 \sigma_2^3+ \sigma_1^2\sigma_2^4
+ \sigma_1\sigma_2^5) + O(\sigma^{10})\nonumber\\
&&= \sigma_1\sigma_2 <({\rm tr} M)^2> + \frac{\sigma_1^5\sigma_2}{5!}<{\rm tr} M^5 {\rm tr} M> + \frac{\sigma_1^4 \sigma_2^2}{4!2!}<{\rm tr} M^4{\rm tr} M^2>\nonumber\\
&& + \frac{\sigma_1^3\sigma_2^3}{3!3!} < ({\rm tr}M^3)^2> + \frac{\sigma_1^2\sigma_2^4}{2!4!} <{\rm tr} M^4{\rm tr} M^2> + \frac{\sigma_1\sigma_2}{5!} <{\rm tr} M{\rm tr} M^5>
+\cdots \nonumber\\
\ea
The coefficient of $\sigma_1^3\sigma_2^3$ gives the knot of $3_1$, which is Seifert graph of $<({\rm tr} M^3)^2>$. Other terms are circle graphs, which are unknotted. From the table A, the knot of two vertices of $U(\sigma_1,\sigma_2)$ are $3_1,5_1,7_1,...$, which are torus knots and they are represented in the replica limit of the average of product of two traces $<({\rm tr} M^{2n+1})^2>$ (n=0,1,2,...).The number of the trace is the number of Seifert disks, and it is equal to  the number of the vertices of Seifert graphs.

For three point vertices, the formula of replica limit (\ref{rf}) gives for the order of $\sigma^{16}$,
\ba\label{3vertex}
&&\lim_{N\to 0} U(\sigma_1,\sigma_2,\sigma_3) |_{\sigma^{16}}= \frac{11}{1152} \sigma_1^8 \sigma_2^4 \sigma_3^4 + \frac{11}{1440}\sigma_1^8 \sigma_2^5 \sigma_3^3 +
\frac{47}{11520} \sigma_1^8\sigma_2^6\sigma_3^2\nonumber\\
&&+ \frac{53}{11520} \sigma_1^7 \sigma_2^7\sigma_3^2+ \frac{347}{34560} \sigma_1^7\sigma_2^6\sigma_3^3+ \frac{89}{5760}\sigma_1^7\sigma_2^5\sigma_3^4+ \frac{623}{34560}\sigma_1^6\sigma_2^6\sigma_3^4\nonumber\\
&&+ \frac{511}{23040}\sigma_1^6\sigma_2^5\sigma_3^5+ {\rm symmetric \hskip 2mm terms}
\ea
which corresponds to the knots of $(8_{8},8_{17},8_{18}, 8_{19},8_{20},8_{21})$ for $\sigma_1^8 \sigma_2^4 \sigma_3^4$. The term of $ \sigma_1^8\sigma_2^5 \sigma_3^3$ corresponds to 
$(8_{16},8_{10},8_7)$. The term of  $\sigma_1^8\sigma_2^6\sigma_3^2$ corresponds to $(8_5,8_2)$ knots.

There is a relation of the knots and the intersection numbers of the curves. The intersection numbers of the curve is a special case of Gromov-Witten invariants which maps from the manifold to a point. The Kontsevich Airy matrix model has an integral representation as \cite{Kontsevich}
\be
Z = \int dM e^{\frac{1}{3}{\rm tr} M^3+ {\rm tr} M^2 \Lambda}
\ee
Expanding the term ${\rm tr} M^3$, we obtain the power of the vertex ${\rm tr} M^3$ for the Gaussian means as $<({\rm tr}M^3)^n>$ (n is even number). We found they correspond to
$3_1$, $9_{23}$, ... knots in the previous section.
The numerical values of the Gaussian means are interpreted as intersection numbers, which are denoted as $<\tau_n>$.
In the case of Kontsevich model, the intersection numbers for one marked point are
\be
<\tau_1> = \frac{1}{24},
\hskip 5mm <\tau_{3g-2}> = \frac{1}{(24)^g g!}
\ee

The Gaussian means are obtained from the derivatives of the logarithm of the partition function $Z$ for the matrix model.

The figure eight $4_1$ knot has an expression of the Gaussian mean $<({\rm tr} M^4) ({\rm tr} M^2)^2>$ in table A.
This is obtained by the derivatives of $c_4$ and $c_2$ for the partition function $Z$, where coefficients of ${\rm tr} M^j$ are $c_j$,
\be
Z = \int dM e^{c_4 {\rm} tr M^4 + c_2 {\rm tr} M^2}
\ee
\be
<({\rm tr} M^4) ({\rm tr} M^2)^2>|_{c_4=0,c_2=-\frac{1}{2}} = \frac{\partial^3}{\partial c_4 \partial^2 c_2} {\rm log Z}
\ee

We consider the k-point function,
\be
U(\sigma_1,...,\sigma_k) = \frac{1}{N} <{\rm tr} e^{\sigma_1 M} \cdots {\rm tr} e^{\sigma_k M}>
\ee
where $M$ is $N\times N$ Hermitian random matrix. There is an integral formula for $k$-point function \cite{BrezinHikami1}
\be\label{N}
U(\sigma_1,...,\sigma_k) = (-1)^{\frac{k(k-1)}{2}} e^{\sum \frac{1}{2}\sigma_i^2} \oint \prod_{i=1} ^k \frac{du_i}{2 i \pi} e^{\sum u_i \sigma_i}
\prod_1^k (1 + \frac{\sigma_i}{u_i})^N {\rm det}\frac{1}{u_i + \sigma_i - u_j}
\ee

In the limit $N\to 0$, we have a logarithmic terms,
\ba
&&\lim_{N\to 0} \frac{1}{N} U(\sigma_1,...,\sigma_k) = (-1)^{\frac{k(k-1)}{2}} e^{\sum \frac{1}{2}\sigma_i^2} \oint \prod_{i=1} ^k \frac{du_i}{2 i \pi} e^{\sum u_i \sigma_i}
\nonumber\\
&&\left(\sum_{i=1}^k {\rm log} ( 1+ \frac{\sigma_i}{u_i })\right) {\rm det}\frac{1}{u_i + \sigma_i - u_j}
\ea
When $k=1$ one point function, this integral is evaluated by the cut of logarithm. The integral of $u$ becomes a line integral $- \sigma <u<0$. (For $k=1$, the determinant is just $\frac{1}{\sigma}$).
\ba\label{onepoint}
&&\lim_{N\to 0} U(\sigma) = \frac{1}{\sigma} e^{\frac{1}{2}\sigma^2} \oint \frac{du}{2i\pi} e^{u \sigma} {\rm log} ( 1+ \frac{\sigma}{u})\nonumber\\
&& = \frac{1}{\sigma} e^{\frac{1}{2}\sigma^2} \int_{-\sigma}^0 du e^{u\sigma} 
=  \frac{{\rm sinh} (\frac{\sigma^2}{2})}{(\frac{\sigma^2}{2})}
\ea
This result is also obtained by the contour integral for the pole $u=0$ by the expansion of the logarithmic term,
\ba
U(\sigma) &=& \frac{1}{\sigma} e^{\frac{\sigma^2}{2}} \oint \frac{du}{2i\pi} e^{u \sigma} (\sum_{n=1}^\infty (-1)^{n-1} \frac{1}{n} (\frac{\sigma}{u})^n)\nonumber\\
&=& 1 + \frac{1}{24} \sigma^4 + \frac{1}{1920} \sigma^8 + O(\sigma^{12})
\ea
which leads to (\ref{onepoint}).

As a categorification of  knot theory, Kontsevich loop expansion has been studied, which provides Vassiliev invariants as coefficients \cite{Kontsevich,Kricker}.  The contribution of one loop order is given by
${{\rm sinh} (\frac{x}{2})}/{(\frac{x}{2})}$, and
the logarithm of this one-loop term  becomes a series expansion of $x^2$ which coefficients $B_n/(4n\cdot (2n)!)$ where  $B_n$ is Bernoulli number. \be
\sum b_{2n} x^{2n} = \frac{1}{2}{\rm log}( \frac{{\rm  sinh}(\frac{x}{2})}{\frac{x}{2}})
\ee
This Knotsevich
one-loop contribution coincides with (\ref{onepoint}). Vassiliev invariant for knot $K$ of order $j$ ( i. e. number of double points in knot $K$) is denoted by $v_j(K)$, and Vassiliev invariants
has a generating function $\sum_j v_j(K) x^j$. This invariants are related to Jones polynomial for knot $K$, $V_K(t)$, with $t= e^{x}$. The expansion of Jones polynomial in power of $x$ 
provides Vassiliev invariants $v_j(K)$ as \cite{Birman}
\be
\sum_j v_j(K) x^j = V_K(e^x)
\ee
The Vassiliev invariant $v_2(K)$ is equal to $(-3)\times $ [$z^2$ coefficient in Conway polynomial $P(z)$]. For instance $v_2(5_2)= -6$ and $P(z)= 1 + 2 z^2$, where $z= \sqrt{t}- \frac{1}{\sqrt{t}}$.
The replica formula of $U(\sigma_1,...,\sigma_n)$, which is used for the intersection theory \cite{BrezinHikami3}, is interesting in the connections of Kontsevich loop expansion and Vassiliev invariants.

For  two point function ($k=2$), we have
\ba\label{N2}
&&\lim_{N\to 0} U(\sigma_1,\sigma_2) = -e^{\frac{1}{2}(\sigma_1^2+\sigma_2^2)}\oint \frac{du_1du_2}{(2i\pi)^2}e^{u_1\sigma_1+u_2\sigma_2}\nonumber\\
&&\times [ ({\rm log}(1+ \frac{\sigma_1}{u_1})+ ({\rm log}(1+ \frac{\sigma_2}{u_2})] \frac{1}{(u_1-u_2+ \sigma_1)(u_2-u_1+\sigma_2)}
\ea
If we consider the contour of $u_1$,  there is no pole for the contour of the logarithmic. Thus it vanishes. Then the integral becomes
the first term,
\ba
&&\lim_{N\to 0} U(\sigma_1,\sigma_2) = -e^{\frac{1}{2}(\sigma_1^2+\sigma_2^2)}\oint \frac{du_1du_2}{(2i\pi)^2}e^{u_1\sigma_1+u_2\sigma_2}\nonumber\\
&&\times  ({\rm log}(1+ \frac{\sigma_1}{u_1}) \frac{1}{(u_1-u_2+ \sigma_1)(u_2-u_1+\sigma_2)}
\ea
By taking the pole of $u_2= u_1-\sigma_2$ and $u_2= u_1+ \sigma_1$, we have
\ba
&&e^{\frac{1}{2}(\sigma_1^2+\sigma_2^2)} \frac{1}{\chi}(e^{-\sigma_2^2}- e^{\sigma_1\sigma_2})\oint \frac{du_1}{2i\pi}e^{u_1\chi } {\rm log} ( 1+ \frac{\sigma_1}{u_1})\nonumber\\
&&=\frac{4}{\chi^2}{\rm sinh} \frac{\chi \sigma_1}{2}{\rm sinh} \frac{\chi\sigma_2}{2}
\ea
where $\chi = \sigma_1+\sigma_2$. Repeating this process, we obtain the replica formula of (\ref{rf}) for $k>2$ \cite{BrezinHikami1}.
 \vskip 3mm
 
  \vskip 3mm
{ \bf{Intersection numbers}}
 \vskip 3mm
 Kontsevich matrix model is given by
 \be
 Z = \int dM e^{\frac{i}{3} {\rm tr} M^3 - \lambda {\rm tr} M^2}
 \ee
 and the intersection numbers for the one marked point $<\tau_n>_g$ is \cite{Kontsevich}
 \be
 <\tau_n>_g = \sum <\tau_n>_g t_n
 \ee
 \vskip 3mm
 The values of the intersection numbers for trivalent matrix model (Airy matrix model) is
 \be
 <\tau_n>_g = \frac{1}{(24)^g g!}
 \ee
 where $n$ is determined by Riemann-Roch relation of Riemann surface,
 \be
 3 g - 2= n
 \ee
 Thus we find the direct relation of trefoil knot to the intersection numbers of $<\tau_1>_{g=1}$. The knots of trivalent Seifert vertices are all alternating knots, and therefore
 the Kontsevich Airy matrix model corresponds to knots $3_1$, $9_{23}$,... one to one. Thus the intersection theory of one marked point gives the knot of trivalent vertices.
 \vskip 3mm
{ \bf{Links}}
 \vskip 3mm
 The $N\to 0$ limit of the matrix expectation values give diagrams  that can be traced on a sheet of paper without lifting a pen : single-stroke diagrams. The next order in $N$ describes two entangled knots, in which the entanglement is  due to a double  layout of two continuous  intricate lines. There again one can obtain the number of entangled knots from the same matrix result.  For the one point function, the linking is obtained from the terms of order $N^2$ of $N\times N$ Hermitian matrix.
 \ba \label{links}
 &&U(\sigma) = \frac{e^{\frac{\sigma^2}{2}}}{N\sigma} \oint \frac{du}{2i \pi} e^{\sigma u} e^{N {\rm log} (1 + \frac{\sigma}{u})}\nonumber\\
 &=&\frac{e^{\frac{\sigma^2}{2}}}{\sigma} \oint \frac{du}{2i \pi} e^{\sigma u} [ {\rm log} (1 + \frac{\sigma}{u}) + \frac{N}{2}
 [{\rm log} (1+ \frac{\sigma}{u})]^2+ O(N^2)
 \ea
 Thus the one-point one link expectations are obtained from the term of order $N$ in (\ref{links}), describing the diagrams made of two strokes trails. The  contour integral in (\ref{links}) provides
 \be U(\sigma) = {e^{\frac{\sigma^2}{2}}}\sum_1  \frac {(N-1)(N-2) \cdots(N-k+1)}{k! (k-1)! }\sigma ^{2k-2} \ee
 and, in the $N\to 0$-limit
 \ba && (N-1)(N-2) \cdots(N-k+1)  = (-1)^{k-1} (k-1)!  \nonumber \\ && +N (-1)^k (k-1)! [ 1+ \frac{1}{2} + \cdots+ \frac{1}{k-1} ] + O(N^2) \ea
 Expressing the sum
 \be  1+ \frac{1}{2} + \cdots+ \frac{1}{k-1}  = \int_0^1 dx \frac {1-x^{k-1}}{1-x} \ee
 one finds the terms of order $N$ in terms of the finite integral \cite{Brezin}
 \ba\label{next}
U(\sigma)|_{N}  &&= \frac{e^{\sigma^2/2} }{\sigma^2}\int_0^1 \frac{dx}{1-x} [   (e^{-\sigma^2 }-1) - \frac{1}{x}(e^{-x\sigma^2 }-1)]\nonumber\\
 && =\frac{1}{2} \sigma^2 +\frac{1}{72} \sigma^6+ O(\sigma^8) \ea
  The one point function $U(\sigma)$ for arbitrary $N$ is expressed as
 \ba
 U(\sigma) &=& \frac{1}{N\sigma} \oint \frac{du}{2 i \pi} e^{\sigma u} (\frac{1+ \frac{\sigma}{2 u}}{1- \frac{\sigma}{2 u}})^N\nonumber\\
 &=& \frac{1}{\sigma} \sum_{k=1}^\infty  \frac{N^{k-1}}{k!} \oint \frac{du}{2i\pi} e^{\sigma u} [{\rm log} (\frac{1+ \frac{\sigma}{2 u}}{1- \frac{\sigma}{2 u}})]^k
 \ea
 For small $\sigma$ expansion becomes
 \ba\label{N0}
 &&U(\sigma) = \sum_{k=1}^\infty \frac{N^{k-1}}{k!}[ \frac{1}{(k-1)!} \sigma^{2k-2} + \frac{1}{12(k+1)!} \sigma^{2k+2} \nonumber\\
 &&+ (\frac{k}{80} 
 + \frac{k(k-1)}{288}) \frac{1}{(k+3)!} \sigma^{2k+6} \nonumber\\
 &&+ (\frac{k}{448} + \frac{k(k-1)}{960} + \frac{k(k-1)(k-2)}{10368}) \frac{1}{(k+5)!}\sigma^{2k+10} \nonumber\\
 &&+ ( \frac{k}{2304} + \frac{71 k (k-1)}{268800} + \frac{k(k-1)(k-2)}{23040} \nonumber\\
 &&+ \frac{k(k-1)(k-2)(k-3)}{497664}) \frac{1}{(k+7)!} \sigma^{2k + 14}+
 O(\sigma^{2k+18} )]
 \ea
 Above expansion is consistent  with (\ref{onepoint}) for $k=1$ and also with (\ref{next}) for $k=2$.
 When the scaling $\sigma^2 \to \sigma^2/N$ is taken, the expansion of the large $N$ is obtained. The leading order of $N$, it is expressed as
 \be
 \lim_{N\to \infty} U(\sigma) = \frac{1}{t}J_1(2t)
 \ee
 where $\sigma= - i t$ and $J_1(x)$ is a Bessel function. The next order of $\frac{1}{N^2}$ is also expressed by the Bessel function of $J_2$. Thus the scaling $\sigma\to \sigma/\sqrt{N}$ in (\ref{N0}) makes  an interesting relation which connects  the large $N$ and the replica limit $N\to 0$.
 
 For the supermatrices in the previous section, one point function is given  with external source eigenvalues $r_i$ and $\rho_j$ \cite{BrezinHikami4s},
 \be
 U(\sigma) = \frac{1}{\sigma}\oint \frac{du}{2i\pi} e^{\sigma u} \prod_{i=1}^n (\frac{u - r_i + \frac{\sigma}{2}}{u - r_i - \frac{\sigma}{2}}) \prod_{j=1}^m (\frac{u - \rho_i - \frac{\sigma}{2}}{u - \rho_j + \frac{\sigma}{2}})
 \ee
 When  the external source $r_i$ and $\rho_j$ are put to zero, this reduces to (\ref{onepoint}) in the limit $n\to m$.
 
 The link graphs are obtained from two point function $u(\sigma_1,\sigma_2)$ in order of $N$.
 For instance, from $<{\rm tr}M^2 {\rm tr} M^2> \sigma_1^2 \sigma_2^2$, with Seifert band connections, there appear a link and two circles, depending the choice
 of crossings. The construction of the Gaussian means are same as knots. There is a correspondence similar to table A between Gaussian means and links.
 The contractions of propagators are given by twisted Seifert bands. For the link of two knots are described by the diagram of order $N$.
 
 \vskip 3mm
 \section{Characteristic polynomial for trivalent vertices}
\vskip 2mm
When Seifert disk (vertex) has trivalent connector, the corresponding Gaussian mean average becomes $<({\rm tr} M^3)^n>$. The non-vanishing one is
the case of $n= 4 m-2, m=1,2,3,...$ from (\ref{select}). The number $m$ is genus $g$ of the Riemann surface. This provides a series of genus $g$ intersection
numbers for trivalent vertices (Kontsevich model) and a series of knots, $3_1, 9_{23},...$.

These knot is classified into 2-bridge knot \cite{Murasugi}, and Seifert matrix $\tilde M$ is made of the diagonal elements plus one line of the same element 1.
From the linking of the Seifert bands of this type of knots, it becomes easy to get Alexander polynomials through the characteristic polynomials,
\be\label{Alexander}
\Delta (t)= {\rm det} ( t \tilde M - \tilde M^T)
\ee
 where $\tilde M$ is Seifert  matrix and $\tilde M^T$ is a transpose of $\tilde M$. For the knot of the series of the trivalent vertices, Seifert matrix $\tilde M$ has diagonal
 elements of 2, except the (1,1) element and (n,n) element, which are 1. For instance, in the case of trefoil $3_1$ and knot of $9_{23}$, Seifert matrix $\tilde M$ become
\be
\tilde M_{3_1}= \left(
 \begin{array}{cc}
1&1\\
0&1
 \end{array}
 \right)
 \ee
\be
\tilde M_{9_{23}}= \left(
 \begin{array}{cccc}
1&1&0&0\\
0&2&1&0\\
0&0&2&1\\
0&0&0&1
 \end{array}
 \right)
 \ee

     \vskip 2mm    
     Alexander polynomial $\Delta(t)$, which is a characteristic polynomial defined by (\ref{Alexander}), has zeros in a complex $t$-plane. In the trivalent vertices of  the previous section,
     the determinant is similar to Toeplitz determinant, and has an interesting zero locus.
     
     It is well known that Ising model with a magnetic field has unit circle zero locus, and there is Yang-Lee edge singularity studied by Fisher \cite{Fisher}. The critical exponent $\sigma$ for  the magnetization $m\sim (h-h_c)^\sigma$ is $\sigma=(d-2+ \eta)/(d+2-\eta)= \frac{\Delta_\phi}{d - \Delta_\phi}$.  The scale dimension $\Delta_\phi= (d-2+\eta)/2$. The exact value of $\Delta_\phi= - \frac{2}{5}$ in Yang-Lee edge singularity is known for two dimensions. 
     This model is given by the Lagrangian ${\mathcal{L}}= \frac{1}{2}(\partial \phi)^2 + i (h-h_c)\phi + i g \phi^3$, and $\epsilon =6-d$ expansion can be obtained.
   The density of characteristic polynomial of trivalent knots have a similarity to edge singularity of $\phi^3$ theory.
     
     The locus of the zero of the characteristic polynomial (Alexander polynomial) is on the arc of the unit circle $x^2+ y^2=1$ in the region $- \sqrt{\frac{3}{4}} < y < \sqrt{\frac{3}{4}}$ and
     $\frac{1}{2}< x <1$ as shown in Fig.1.  The density of the zeros increases in the approaching to the edge point in the complex plane $(x,y)= (\frac{1}{2}, \sqrt{\frac{3}{4}})$. This point is analogous to the
     edge point of the zeros of Yang-Lee Ising model in a magnetic field. The distribution of zeros for trivalent vertex knot shows the square root singularity as the edge singularity. This behavior is originated from the relation of cusp singularity $y^2= x^3$ to the trefoil knot. This is also related to Landau- Ginzburg potential for primary fields, which are
      obtained from matrix models  \cite{Witten1,BrezinHikami4}.

     The knot $5_1$ has a vertex with 5 external legs. This is related to $p$-spin curve with $p=4$.  In this case, the same ladder structure for higher crossing knots made of   vertices with 5-legs only as expressed as $<(\rm tr M^5)^n>$, $n= 3, 6,...$.
     The zeros of Alexander polynomials are located on the unit circle similar to the trivalent vertices. The region of the zeros on the unit circle is limited to the region $(-1)^{\frac{3}{5}} < x$, where $x$ is the real part of the zeros. The zeros at $(-1)^{\frac{1}{5}}$ and $(-1)^{\frac{3}{5}}$, for positive imaginary region, appear independent of $n$, and they are the end points
     of edges of a spectrum.


 
     \begin{figure}[htbp]
\begin{center}
\includegraphics[width=0.3\textwidth]{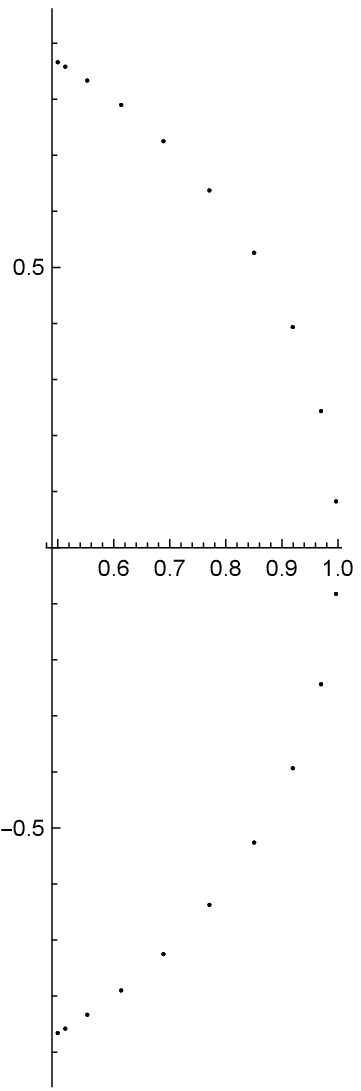}
\caption{\small{Zeros of Alexander polynomial for trivalent vertex, which are on an arc of the unit circle.
The edge points of spectrum are at $(x,y)= (\frac{1}{2}, \pm\sqrt{\frac{3}{4}})$ in a complex coordinate.  }}

\end{center}
\end{figure}
     \vskip 50mm

   \vskip 3mm
  \section{Higher dimensional knots and singularities}
     \vskip 3mm  
     The classical knot theory concerns with the embedding of $S^1$ into $S^3$.   The extension of this knot to higher dimensional case has been studied with covering spaces of the classical knots \cite{Murasugi, Kamada}.
      $n$-dimensional knot  $(N,M,k)$ is embedding of $n$-dimensional manifold $N^n$ in  $(n+2)$ dimensional manifold $M^{n+2}$, ($k : N^n \subset M^{n+2}$). The embedding $k$:$N^n\subset M^m$ with codimension $m-n\ge 3$ leads to only  unknotting.  So we need the codimension 2 for knots.
  A classical knot is 1-knot, $n=1$. By spinning classical knots,  the surface knot (2-knots) is obtained, which is embedded in 4-manifold ($S^2\subset S^4$)).    
   
   For 2-knot, the Seifert surface is constructed by the spinning of a knot around a certain axis, which leads to generation of surfaces \cite{Zeeman}.    
Natural way of 2-knot is obtained from 1-knot by taking the movement in a time direction, i.e. movie picture \cite{Kamada}.
    
    The replica limit of (\ref{rf}) is applied for  the time dependent Gaussian matrix model \cite{BrezinHikami4}.  By adding ${\rm tr} (d M /dt)^2$ term to action, the matrix describes the movement  on time.  This time dependent
 matrix model reduces to two matrix model by  a path integral formulation \cite{BrezinHikami4}. 
 
   When 1-knot is embedded in to 4 dimensions, the knot becomes resolved due to the additional dimension of
 time. However, the pair of strings, which makes a surface becomes knotted. The axis of the spun knot correspond the interaction of two matrices $M_1$ and $M_2$ coupled as $(c {\rm tr} M_1 M_2)^2$.
 
 The 2-knot has  singularities such as double point and triple point, which are generated by the degeneracies of Reidemeister relation in the moving picture \cite{Kamada}. The Reidemeister relations for 1-knot are generalized to Roseman relations. Over and down distinction in 1-knot with previous numberings does not make a sense for 2-knot. The coloring (numbering) may be useful instead \cite{Saito}. We consider the replica limit of the time dependent matrix model for the description of the surface knot.
 
  The two point correlation function $U(\sigma_1,\sigma_2)$ is
 \be\label{two}
 U(\sigma_1,\sigma_2)= \frac{1}{Z} \int dM_1 dM_2 ({\rm tr} e^{\sigma_1 M_1} {\rm tr} e^{\sigma_2 M_2}) e^{-\frac{1}{2} {\rm tr} (M_1^2 + M_2^2 - 2 c M_1 M_2)}
 \ee
 The parameter $c$ is a coupling constant of two matrices, which is equal to  $e^{-t}$ with  a time $t$. 
 In the limit $t \to \infty$, $c$ becomes vanishing, and we obtain two non-interacting matrices.  
 
 For the picture of the spinning, we take order the terms of $c^2$, which represent the axis of the spinning. 
 The n-point correlation function for $c\ne 0$ can be formulated in a contour integrals \cite{BrezinHikami4}.
    The replica limit for this case is obtained by taking $N \to 0$ in the contour integral. For instance, the two point case is expressed as
    \ba\label{twomatrix}
    U(\sigma_1,\sigma_2) &=& - \frac{1}{N} \oint \frac{du dv}{(2 i \pi)^2} ( 1+ \frac{\sigma_1}{u})^N (1+ \frac{\sigma_2}{c v})^N \nonumber\\
    &\times& \frac{1}{(u-v+ \frac{\sigma_1}{N})(u- v + \frac{\sigma_2}{N})} 
    e^{\frac{\sigma_1u}{1- c^2} + \frac{\sigma_2 v}{1-c^2}- \frac{\sigma_1^2}{2N(1-c^2)} + \frac{\sigma_2^2}{2 N (1- c^2)}}
    \ea

  \vskip 3mm
  \section{Extension to half integer spins}
     \vskip 3mm  
     Recently the extension of the $p$ spin curve of integral value of $p$ to half integer $p$ has been considered by Br\'ezin and myself  \cite{BrezinHikami5,BrezinHikami6,Hikami7}.
     The skein relation is expressed with a parameter  $z= \sqrt{t}- \frac{1}{\sqrt{t}}$.  This parameter $t$ appears in Alexander polynomials and Jones polynomials \cite{Jones}. This parameter $t$ can be a root of unity,
     $t = e^{\frac{2 \pi i}{p}}$. When $p$ is half-integer, the skein relation provides a new relation of knots. There appears also a factor $\tilde z= \sqrt{t}+\frac{1}{\sqrt{t}}$, which represents
     a trivial link . It is interesting to note that this factor appears in the contour integral with a change variable from $u$ to $y$ as $u= \frac{i}{2}(y^2-\frac{1}{y^2})$, which gives a Laurent series in $y$ \cite{BrezinHikami5,BrezinHikami6,Hikami7}. In the expression of the integrand for correlation function, there appears a factor $\tilde z =\sqrt{t}+ \frac{1}{\sqrt{t}}$ with $y= \sqrt{t}$. The $\mu$ component trivial link is expressed as $V_{O_\mu}(t) = (-1)^{\mu-1}(\sqrt{t} + \frac{1}{\sqrt{t}})^{\mu-1}$, which appears in the construction of Jones polynomials \cite{Murasugi}.
     
     This half-integer spin leads to Fermionic case. The case $p= \frac{1}{2}$ is Dirac spin and $p = \frac{3}{2}$ is Rarita-Schwinger case. Such half integer $p$ and half integer level $k$, ($p= k+ 2$) has a new properties for knots. Since $k$ is a coefficient of Chern-Simons action, this fractional level leads to a fractional charge and a new conformal field theory, which may have application on   topological semi-metals of spin $\frac{3}{2}$ such as Half-Heusler alloy (PdBiSe etc.).

\section{Discussion}
\vskip 2mm
 In this article, we point out explicitly that the expectation values of vertices for the Gaussian Hermitian matrix model, corresponds to knots  in the zero-replica limit.  The diagrammatic expansions
show an  explicit correspondence between Seifert graphs and  Gaussian means after introduction of an height function. By the tuning of an external matrix source,
 the Hermitian matrix model can be chosen to generate $p$-spin curves \cite{BrezinHikami3}. We found the correspondence of the $n$-vertex Gaussian means  to the singularity
 of Landau-Ginzburg potentials. This correspondence appears in the distribution of zeros of the corresponding Alexander polynomial, which is related to Yang-Lee edge singularity.
  
 The replica limit of $O(N)$ vector model has been used for polymers as a way to implement the self-avoiding behavior in $d$ dimensions, where $1<d<4$. Recently, polymers have been studied by a conformal bootstrap method
 for general space dimensions $d$ \cite{ShimadaHikami,Hikami03}. The Yang-Lee edge singularity has been studied by the conformal bootstrap in a  determinant method \cite{Gliozzi,Hikami04} for dimensions $2 \le d \le 6$. The value  $\Delta_\phi \sim -\frac{2}{5}$ has been found for two dimensions. The $d$ dimensional branched polymer is equivalent to Yang-Lee edge singularity with the dimensional reduction $d \to d-2$ , which can be seen explicitly in  Wilson-Fisher $\epsilon$ expansion \cite{WilsonFisher,Hikami03}.
It is interesting to investigate the dimensional reduction \cite{Hikami05} for the knots of higher dimensions. This  will be  an interesting  future work.

 Recently, in a four-dimensional ${\mathcal{N}} = 4$ super Yang-Mills theory, similar replica formula of (\ref{rf}) appears (see footnote 15 in ref. \cite{Vieira}). This may be an indication that a gauge theory appears from the random matrix theory with a replica.
 
 The spin glass in a disordered system has been discussed by a replica method as a replica symmetry breaking \cite{Parisi}, which may be equivalent to $p$-adic theory \cite{Parisi2}.  We have 
 seen the knot or unknotted circle appear in the random matrix theory with replica limit. The unknotted circles provide a large entropy, and a glass transition may be considered as a 
 transition from
 the knotted system to unknotted (circle) transitions. It may be interesting to apply the present study of the replica limit of the matrix model to such glass transition and gauge glass transition as a vortex lattice
 melting in a superconductor \cite{BrezinHikamiFujita,Fujita}.

     \vskip 5mm
      {\bf Acknowledgement}
      \vskip 3mm
      Author thanks  Edouard Br\'ezin for many discussions about replica method and fractional $p$ spin curves.
      He  thanks Andreani Petrou for the help of evaluation of knots. He is supported by JSPS KAKENHI 19H01813.

\end{document}